\documentstyle[aps,prl,twocolumn,epsfig,rotate]{revtex}  
\tighten

\def\DEDX{dE/d{\it x }}
\def\dplus{${D^{+}}$}
\def\dminus{${D^{-}}$}
\def\dzero{${D^{0}}$}
\def\dzerobar{${\overline{D}^{0}}$}
\def\dstarp{${D^{*}(2010)^{+}}$}
\def\dstarm{${D^{*}(2010)^{-}}$}
\def\dstarz{${D^*(2007)^0}$}
\def\dstarzb{${\overline{D}^{*}(2007)^0}$}

\newcommand{\epem}   {$e^{+}e^{-}$}

\newcommand{\rarr}     {\rightarrow}

\title{Direct Measurement of B$({D^0\rightarrow\phi X^0})$ and
 B$({D^+\rightarrow\phi X^+})$ } 

\author{
J.~Z.~Bai,$^1$   Y.~Ban,$^5$      J.~G.~Bian,$^1$
I.~Blum,$^{12}$ 
G.~P.~Chen,$^1$  H.~F.~Chen,$^{11}$  
J.~Chen,$^3$ 
J.~C.~Chen,$^1$  Y.~Chen,$^1$ Y.~B.~Chen,$^1$
% ...
\cite{atNU0}
Y.~Y.~Wang,$^1$  
M.~Weaver,$^2$
C.~L.~Wei,$^1$   N.~Wu,$^1$       Y.~G.~Wu,$^1$
D.~M.~Xi,$^1$    X.~M.~Xia,$^1$   P.~P.~Xie,$^1$    Y.~Xie,$^1$
Y.~H.~Xie,$^1$   G.~F.~Xu,$^1$    S.~T.~Xue,$^1$
J.~Yan,$^1$      W.~G.~Yan,$^1$   C.~M.~Yang,$^1$   C.~Y.~Yang,$^1$
H.~X.~Yang,$^1$  J.~Yang,$^1$     
W.~Yang,$^3$
X.~F.~Yang,$^1$  M.~H.~Ye,$^1$    S.~W.~Ye,$^{11}$
Y.~X.~Ye,$^{11}$   C.~S.~Yu,$^1$    C.~X.~Yu,$^1$     G.~W.~Yu,$^1$
Y.~H.~Yu,$^4$    Z.~Q.~Yu,$^1$    C.~Z.~Yuan,$^1$   Y.~Yuan,$^1$
B.~Y.~Zhang,$^1$  C.~Zhang,$^1$   C.~C.~Zhang,$^1$  D.~H.~Zhang,$^1$  
Dehong~Zhang,$^1$
H.~L.~Zhang,$^1$ J.~Zhang,$^1$    J.~W.~Zhang,$^1$  L.~Zhang,$^1$
L.~S.~Zhang,$^1$ P.~Zhang,$^1$
Q.~J.~Zhang,$^1$ S.~Q.~Zhang,$^1$ X.~Y.~Zhang,$^6$  Y.~Y.~Zhang,$^1$
D.~X.~Zhao,$^1$  H.~W.~Zhao,$^1$  Jiawei~Zhao,$^{11}$ J.~W.~Zhao,$^1$
M.~Zhao,$^1$     W.~R.~Zhao,$^1$  Z.~G.~Zhao,$^1$   J.~P.~Zheng,$^1$
L.~S.~Zheng,$^1$ Z.~P.~Zheng,$^1$ B.~Q.~Zhou,$^1$   G.~P.~Zhou,$^1$
H.~S.~Zhou,$^1$  L.~Zhou,$^1$     K.~J.~Zhu,$^1$    Q.~M.~Zhu,$^1$
Y.~C.~Zhu,$^1$   Y.~S.~Zhu,$^1$   B.~A.~Zhuang$^1$
\\ (BES Collaboration)}

\address{
$^1$Institute of High Energy Physics, Beijing 100039, People's Republic of
 China\\
$^2$California Institute of Technology, Pasadena, California 91125\\
$^3$Colorado State University, Fort Collins, Colorado 80523\\
$^4$Hangzhou University, Hangzhou 310028, People's Republic of China\\
$^5$Peking University, Beijing 100871, People's Republic of China\\
$^6$Shandong University, Jinan 250100, People's Republic of China\\
$^7$Shanghai Jiaotong University, Shanghai 200030, People's Republic of China\\
$^8$Stanford Linear Accelerator Center, Stanford, California 94309\\
$^9$University of Hawaii, Honolulu, Hawaii 96822\\
$^{10}$University of California at Irvine, Irvine, California 92717\\
$^{11}$University of Science and Technology of China, Hefei 230026,
People's Republic of China\\
$^{12}$University of Texas at Dallas, Richardson, Texas 75083-0688}

\date{\today}
\begin{document}
\maketitle

\begin{abstract}  
The first measurement of B(${D^0\rightarrow\phi X^0}$)
and an upper limit for
B(${D^+\rightarrow\phi X^+}$)
are determined from 22.3 {\it pb$^{-1}$} of
\epem\ annihilation data at a C. M. energy of 4.03 GeV.
The data were recorded by the Beijing Spectrometer
(BES) at BEPC. A recoil charge method is applied for the first
time to charm threshold data to determine the charge of the 
{\it D} meson in the recoil from 
$9054\pm309\pm416$ reconstructed \dzero , \dplus\ mesons.
The branching fractions 
$B({D^0\rightarrow\phi X^0}) =(1.71^{+0.76}_{-0.71}\pm0.17)$\%,
and $B({D^+\rightarrow\phi X^+}) <1.8$\% are
determined from 10 events with a reconstructed {\it D}
and a recoiling $\phi$.  
In addition, a 
$90\%$ C.L. upper limit of
B$({ D^+\rightarrow \phi e^+X^0})<1.6\%$ is determined
from a search for semileptonic decays of the \dplus .

\end{abstract}

\section{Introduction}

Much progress has been made in heavy flavor physics since the
discoveries of open charm and bottom\cite{cb_discovery}.
However, experimental results on inclusive
decays of charm mesons are limited\cite{drev,pdg}.
In an era of high precision
experiments such as the $B$ factories and the LHC, accurate measurements
of {\it b}-flavored particles can benefit from 
a better knowledge of charm decays and their branching
fractions.
In particular,
the inclusive decay 
${D\rightarrow\phi X}$ \footnote{
Throughout this paper, charge conjugation invariance is assumed, and
charge conjugate states are included.}
has not been measured.
This branching fraction serves as an
independent check for the existence of
additional exclusive decays of 
{\it D} mesons that contain a $\phi$
meson. It is 
important for studies of gluonic penguin ($b \rarr sg$)
decays of {\it B} mesons\cite{pdg},
and for time-dependent ${B^0_s\overline{B}^0_s}$
oscillation measurements\cite{LEP_bs_mixing} 
that use a $\phi\ell$ pair to tag ${B_s^0}$ mesons.
In addition, this branching fraction would be helpful in understanding
the charm meson decay mechanisms.

In this letter, we report a first measurement of the inclusive $\phi$
decay branching fractions of charged and neutral {\it D} mesons and a 
search for the exclusive semileptonic decay
${D^+ \rarr \phi e^+X^0}$ 
based on an analysis of 
22.3 {\it pb$^{-1}$} of 
data collected with the Beijing Spectrometer (BES)
in $e^+e^-$ annihilations at $\sqrt{s}=4.03$ GeV.

\section{The BES Detector}
The BES detector has been described in detail elsewhere\cite{bes}.
Here we briefly
describe detector elements crucial to this measurement.

BES is a conventional solenoidal detector
operated at the Beijing Electron Positron Collider (BEPC)\cite{bepc}.
A four-layer central drift
chamber (CDC) surrounding the beam pipe 
is used for triggering purposes.
A forty-layer main drift chamber (MDC) located just outside the CDC provides
measurements of charged tracks and 
ionization energy loss (\DEDX)
with a solid angle coverage of $80\%$ of $4\pi$ for charged tracks.
A momentum resolution of $1.7\%\sqrt{1+p^2}$ ($p$ in {GeV}/$c$)
and a \DEDX resolution of $8.5\%$ for Bhabha electrons 
are obtained for 
data taken at $\sqrt{s}=4.03$ GeV. An array of 48 scintillation counters
surrounds the MDC and measures the time of flight (TOF) of charged tracks
with a resolution of about 350 ps for Bhabha electrons and 450 ps for
hadrons.
Surrounding the TOF is a 12-radiation-length, lead-gas barrel shower counter
(BSC) operated in limited streamer mode, which
measures the energy of electrons
and photons over $80\%$ of 4$\pi$, with an energy
resolution of $\sigma_E/E=0.22/\sqrt{E}$($E$ in GeV), and 
spatial resolution of $\sigma_{\phi}=4.5$ mrad and $\sigma_Z=2$ cm for 
electrons. Outside the BSC is a solenoidal magnet providing a
0.4 T magnetic field for the central tracking region of the detector. Three
double-layers of muon counters instrument the magnet flux return, and serve to
identify muons with transverse momenta 
greater than 500 MeV/$c$. They cover $68\%$ of 
4$\pi$ with a 
longitudinal (transverse) spatial resolution of 5 cm (3
cm). 

\section{The recoil charge method for $D$ type identification}

At $\sqrt{s}$=4.03 GeV charm mesons $D^0$ and $D^+$ 
are produced via the interactions\\
\indent
{\hskip 1.5cm}
${e^+e^-}  \rarr { D^+D^-,D^0\overline{D}^{0}}, $\\
\indent
{\hskip 2.9cm} \dstarp\dminus ,  \dstarz\dzerobar ,\\
\indent
{\hskip 2.9cm} \dstarp\dstarm ,\\
\indent
{\hskip 2.9cm} \dstarz\dstarzb \\

\noindent and possibly $D\overline{D}\pi$. 
The \dstarm\ can decay to either $\pi^{-} \overline{D}^{0}$
or $\pi^0(\gamma) {D^{-}}$. The prompt or transition
pion is not identified.
Reconstructing a specific $D$ meson does not necessarily determine
whether the recoiling 
${\overline{D}}$  meson is charged or neutral.
In order to measure 
$B{(D^0\rightarrow \phi X^0)}$ and
$B{(D^+\rightarrow \phi X^+)}$ specifically,
the numbers of neutral and 
charged ${\overline{D}}$ mesons recoiling against a reconstructed $D$ meson,
which tags the ${e^+e^-\rightarrow D\overline{D}}$ events,
and the types of the $D$ meson from 
which the $\phi$ mesons come,
must be determined.

 \begin{figure}
 \center{\mbox{\epsfig{file=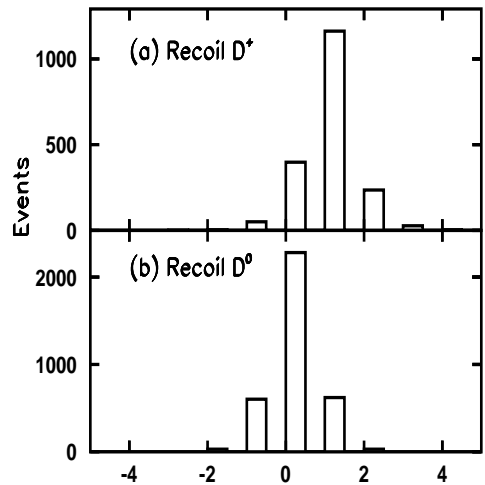,height=8.0cm,width=8.0cm}}}
 \vspace*{-0.2cm}
 \caption{
    Monte Carlo $Q_{rec}$ distributions for
    (a)  $D^+$ decays, and (b)  $D^0$ decays.
    In both cases the $D$ mesons are in the recoil
    against a
    fully reconstructed charm meson.
}
\end{figure}

A recoil charge method for identifying the type of the recoiling $D$ meson
has been devised for this measurement.
At $\sqrt{s}=$4.03 GeV, 
charged pions arising from \dstarp\ decay are very slow, and are
mostly undetected in the BES detector. Only
charged particles from decays of
${D^+}$ and ${D^0}$ are measured,
and their total charge is correlated with the type of 
the mother $D$ meson.
Figures 1(a) and 1(b) show the
Monte Carlo distributions of total
detected recoil charge, $Q_{rec}$, of  
$D^+$ and $D^0$ mesons against a fully
reconstructed $\overline{D}$ meson, respectively.
They are centered at +1 and 0, respectively, and have a
spread of about +1 or -1. 
The recoil charge method selects
neutral and charged $D$ mesons according to

\begin{equation}
{ Q_{rec}=0, ~~{\rm or}~~Q_{rec}=F_{D}=-1 ~{\rm for}~~{D^0}}
\end{equation}

\noindent and
\begin{equation}
{Q_{rec}\cdot F_{D}<0~~{\rm for}~~ {D^+}}
\end{equation}
\noindent
where $F_{D}$ is the charm quantum number 
of the reconstructed $\overline{D}$ meson. 
The efficiency,  $\varepsilon$, and the misidentification rate, 
$f$, of
the recoil charge method are 
obtained from Monte Carlo simulations. For inclusive
$D$ decays, $\varepsilon$ and $f$ are determined to be
0.74$\pm$0.02 and 0.25$\pm$0.02, respectively, and 
are the same for both charged and neutral 
recoiling {\it D} mesons.
These values are confirmed with data using
a  kinematically selected data sample of  
${e^+e^-\rightarrow D^+D^-}$
and ${e^+e^-\rightarrow D^0\overline{D}^{0}}$,
where  one {\it D} meson has been reconstructed.
The values from data,
$\varepsilon$=0.79$\pm$0.04, $f=0.16\pm$0.04 
agree well with the Monte Carlo estimates of
$\varepsilon$=0.795$\pm$0.015, 
$f=0.204\pm$0.015 for the exclusive interaction 
${e^+e^-\rightarrow D\overline {D}}$.
The slightly higher efficiency for this
sample is due to 
the absence of charged transition pions. 

When both a 
{\it D} meson and a recoil $\phi$ are fully reconstructed,
the efficiency of the recoil charge method is improved over
that of the inclusive {\it D} events because 
there are fewer remaining tracks 
and they tend to cluster  
along the $\phi$ direction and are therefore well 
within the detector acceptance.
A Monte Carlo study of various
{\it D} decay modes into final states containing
a $\phi$ has been
performed, and the variations among their efficiencies
are included in the systematic errors. 
For events with a reconstructed 
$\phi$, the recoil charge method 
identifies the recoil {\it D} meson type correctly 
(91$\pm$1$\pm$2)\% of the time and
misidentifies a {\it D} in
(9$\pm$1$\pm$2)\% of the events.
The first error
is due to Monte Carlo statistics, and the second is 
systematic.

To determine the numbers of neutral and charged $D$ mesons
on the recoil side, we use the relationships

\begin{equation}
{ \left( \begin{array}{c}
N_{0} \\
N_{+}
\end{array} \right) = \left( \begin{array}{cc}
\varepsilon & f \\
f & \varepsilon
\end{array} \right)  \left( \begin{array}{c}
N_{D^0} \\
N_{D^+}
\end{array} \right) \label{eq:eq1} }
\end{equation}

\noindent and

\begin{equation}
{r_{+}={ {N_{D^+}}\over {N_{D^0} + N_{D^+}}}
={ {\varepsilon N_{+} - fN_{0}}\over
{(N_{D^0} + N_{D^+})(\varepsilon^2 - f^2)} } \label{eq:eq2} }
\end{equation} 

\noindent
where $N_{0}$ and $N_{+}$ are the events that pass the recoil charge
tag as neutral and charged {\it D} meson candidates, respectively, 
$N_{D^0}$ and $N_{D^+}$
are the true numbers of neutral and charged 
{\it D} mesons in the recoil,
and $r_{+}$ is the fraction of recoiling $D^+$
in the {\it D} tag sample.
The values of $N_{0}$ and $N_{+}$ are obtained directly from data.
Using Eqs. (3) and (4) 
$N_{D^0}$ and $N_{D^+}$ in each
{\it D} decay mode are determined.

\section{Data Analysis}

\subsection{Event selection}

Charged tracks are required to have good
helix fits.
These tracks must satisfy $|\cos\theta|<$ 0.8,
where $\theta$ is the polar angle, 
and must be consistent with originating from the primary
event vertex.
For charged particles, 
a combined confidence level 
calculated using the \DEDX  and TOF 
measurements is required to be greater
than $1\%$ for the $\pi$ hypothesis.
For the kaon hypothesis, ${\it L_k>L_{\pi}}$ is required,
where ${\it L}$ is the likelihood for a 
particle type.

\subsection{Analysis of inclusive $D$ meson events}

Charged and neutral $D$ mesons are reconstructed in the 
${D^0 \rarr K^{-}\pi^{+}}$,
${K^{-}\pi^{-}\pi^{+}\pi^{+}}$ and
${D^{+} \rarr K^{-}\pi^{+}\pi^{+}}$ decay modes.
Figures 2(a), 2(b) and 2(c) show the momentum distributions of
selected ${Kn\pi}$ combinations
with invariant masses within
$\pm$2.5 standard deviations of the nominal
$D^0$ and $D^+$ masses.
The lowest momentum peak in each figure
corresponds to $D$ mesons from
${e^+e^-\rightarrow D^*\overline{D}^{*}}$, the middle peak is 
predominantly due to
${e^+e^-\rightarrow D\overline{D}^{*}}$, and
the small enhancements at high momentum are from
direct ${e^+e^-\rightarrow D\overline {D}}$ production.

\begin{figure}
\center{\mbox{\epsfig{file=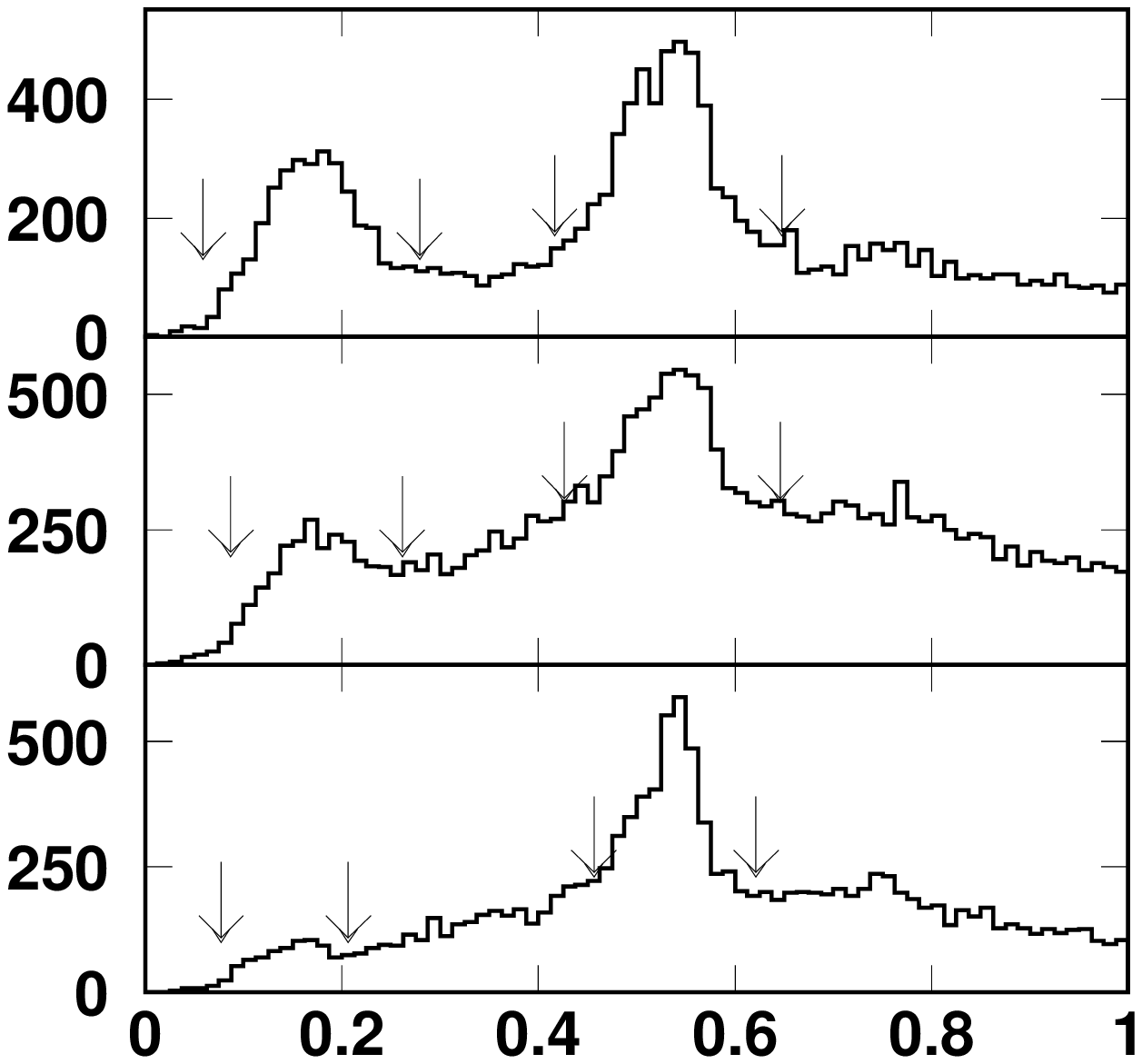,height=9.0cm,width=8.0cm}}}
\put(-165,8){\bf\large Momentum (GeV/$c$)}
\put(-240,80){\bf\rotate{\large Events / 12.5 MeV/$c$}}
\put(-60,200){(a)}
\put(-60,140){(b)}
\put(-60,80){(c)}
\vspace*{0.1cm}
\caption{Momentum distribution of (a) the ${K^-\pi^+}$,
(b) the ${K^-\pi^-\pi^+\pi^+}$, and
(c) the ${K^-\pi^+\pi^+}$
combinations that pass the selection
criteria and the mass cuts mentioned in the text.}
\end{figure}

To reduce combinatorial backgrounds 
only {\it D} mesons in the lower two momentum regions 
are used in this measurement. 
Figures 3(a), 3(b) and 3(c) show the invariant mass
distributions for selected ${Kn\pi}$ pairs.
A binned maximum-likelihood fit to the distributions
with a Gaussian signal function and a third order polynomial
background
yields
$10371\pm357$ events.
Some $D$ events enter the $D$ signal region more than once
due to $K-\pi$ interchange and random combinations,
and are doubly counted in the fit.
The rates of double counting are evaluated to be 
14\% for $D^0$ and 5\% for $D^+$, using 
data and Monte Carlo events. The largest variation
in these rates
between data and Monte Carlo simulation is found to be
3.6\% and is included in the systematic error.
Specific to the ${K^-\pi^+\pi^+\pi^-}$ mode, 
partially reconstructed $D^0$ events combine with
slow pions to produce an enhancement at the $D^0$ signal
position with a resolution similar to that of the signal.
A Monte Carlo simulation indicates this enhancement
accounts for as much as 6.2\% of the number of
${D^0 \rightarrow K^-\pi^+\pi^+\pi^-}$ events
given by the fit. This variation is also
included in the systematic error.

\begin{figure}
\center{\mbox{\epsfig{file=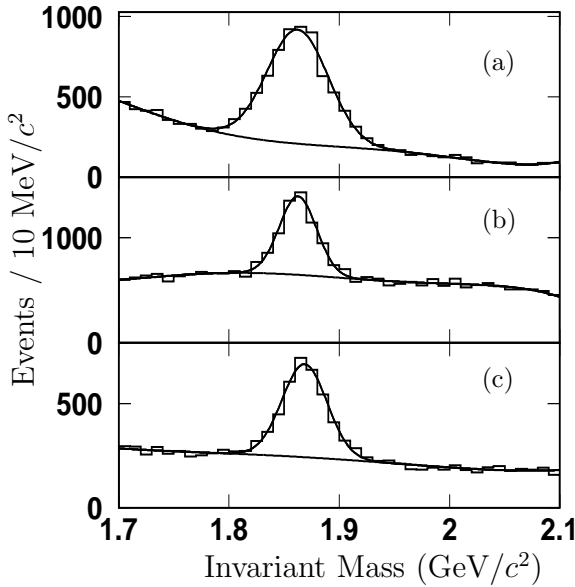,height=9.0cm,width=8.0cm}}}
\put(-165,8){\bf\large Invariant Mass (GeV/$c^2$)}
\put(-239,80){\bf\rotate{\large Events / 10 MeV/$c^2$}}
\put(-60,200){(a)}
\put(-60,140){(b)}
\put(-60,80){(c)}
\vspace*{0.1cm}
\caption{Invariant mass distribution of selected
(a) $K^{-}\pi^{+}$,
(b) $K^{-}\pi^{+}\pi^{+}\pi^{-}$,
and (c) $K^{-}\pi^{+}\pi^{+}$ combinations.}
\end{figure}

After subtracting the doubly counted events, a sample of
$9054\pm309\pm416$ tagged events is selected which contains 
in the recoil 
6767$\pm$298$\pm$446 \dzero\ 
and 2287$\pm$83$\pm$155 \dplus\ 
mesons, as determined by the recoil
charge method using Eqs (3) and (4), where the first errors
are statistical and the second systematic. 
The systematic errors are due to
variations in
the mass fits, the subtraction of doubly counted events,
and the uncertainty of the efficiency of the recoil charge
method.
A summary of these events
is presented in Table I. 
These recoiling $D^0$ and $D^+$ mesons are
unbiased {\it D} decay samples for measuring
branching fractions.

\begin{small}
\begin{table}
\centering
\vspace{0.2cm}
\caption{Determination of $D$ meson sample}
\begin{tabular}{|c|c|c|c|c|}
$D$ tag   & number of & $r_+$ & $N_{D^0}$ & $N_{D^+}$ \\
type    & $D$ events  &       &               &                \\
\hline
\dzero\ & 6895$\pm$288&
0.20 & $5513\pm 280$ & $1382 \pm 69$ \\
\hline
\dplus\    &2159$\pm$112 &
  0.42 & $1254 \pm 102$ & 905 $\pm$ 47 \\
\hline
Sum      & 9054$\pm$309   &  &
$ 6767\pm 298 $ & $ 2287\pm 83  $ \\
\end{tabular}
\end{table}
\end{small}

An independent estimate of the number of recoiling
\dzero\ and \dplus\ has been carried out using
measured production cross sections of
${D^*\overline{D}}$ and
${D^*\overline{D}^{*}}$ and the branching fractions
of ${D^*}$ and $D$ mesons.
The result is consistent with the recoil charge
method.

\subsection{Measurement of branching fractions}

\subsubsection{Reconstruction of ${\phi\rightarrow K^+K^-}$ events}

The $\phi$ meson is reconstructed 
through its decay to ${K^+K^-}$.
Figure 4 shows the invariant mass distribution of ${K^+K^-}$
pairs.
A fit of convoluted Breit-Wigner and Gaussian functions plus a third order
polynomial background gives
a fit mass of $1.0194\pm0.0002$ GeV$/c^2$ and a total of 
$1108\pm 70$ $\phi$ events. In this measurement, the $\phi$ signal window
is defined as the region from 1.00 to 1.04 GeV/$c^2$, as indicated by the arrows
in Figure 4. 

\begin{figure}
\center{\mbox{\psfig{figure=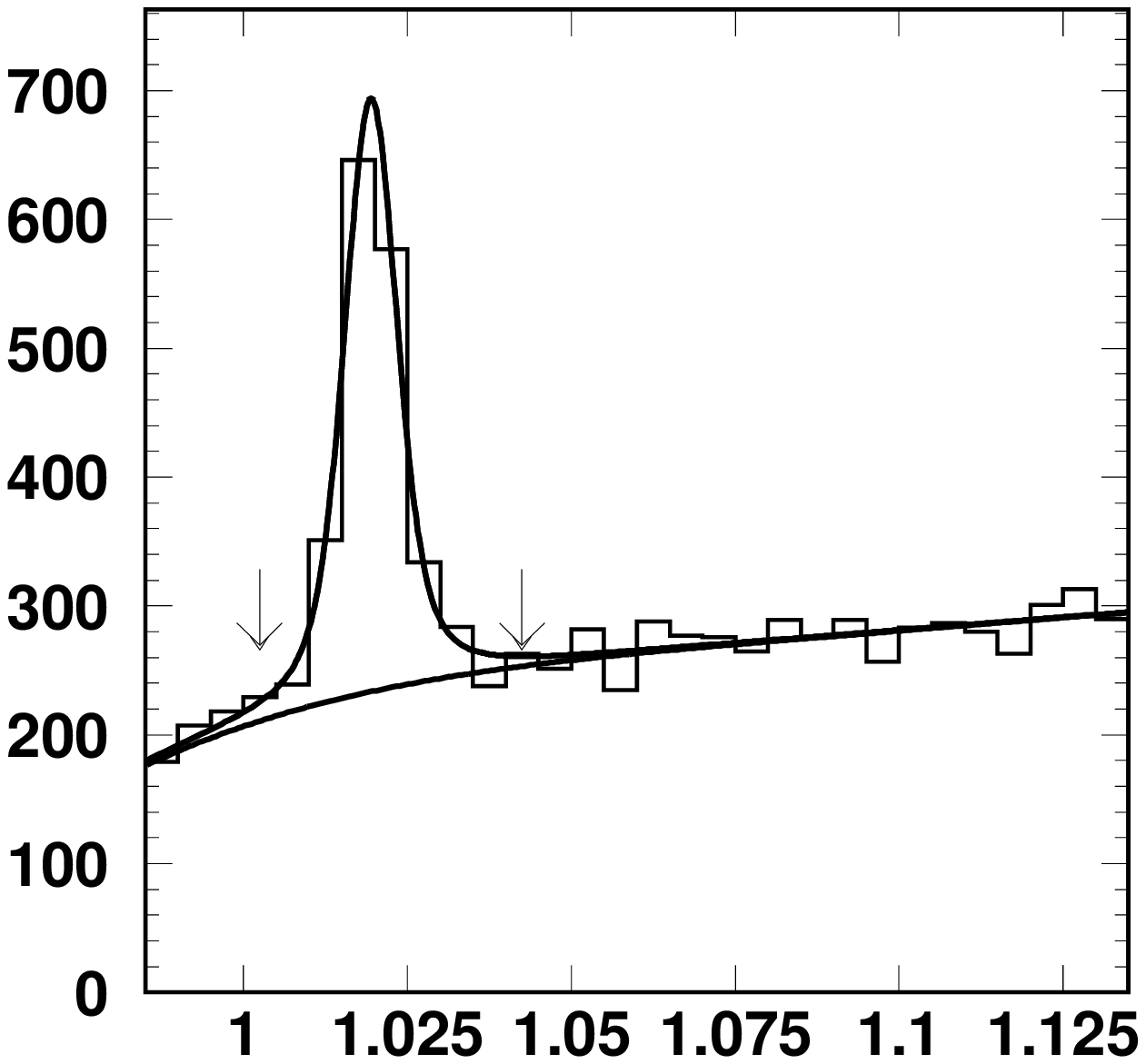,height=7.0cm,width=8.0cm}}}
\put(-165,5){\bf\large Invariant Mass (GeV/$c^2$)}
\put(-235,60){\bf\rotate{\large Events / 5 MeV/$c^2$}}
\vspace*{0.1cm}
\caption{Invariant mass of inclusive ${K^+K^-}$ pairs.}
\end{figure}

\subsubsection{Inclusive ${D\rightarrow \phi X}$ }

Figures 5(a) and 5(b) 
show the mass of ${K^+K^-}$ pairs recoiling against
\dzero\ and \dplus\ candidates, respectively,
and the full data are shown in Figure 5(c).

To estimate the number of signal events, the ${K^+K^-}$
mass 
intervals 0.98-1.00 GeV/$c^2$ and 1.04-1.15 GeV/$c^2$ are
taken as background regions for the $\phi$.
The ${Kn\pi}$ mass
regions from 1.7 to 2.1 GeV/$c^2$, excluding
regions within ${\pm3\sigma_{M_D}}$ of the fit
$D$ masses,
are defined as sideband background control regions for the
$D$ mesons. As shown in Figure 5(c), 15 events are found as
${D\phi}$ candidates, and 14 events are selected
as background sample outside the $\phi$ mass region.
Using the $D$
side band events, a total of 0.5$\pm$0.5 events
has been estimated as the background among the $D$ candidates.
Subtracting the background contributions to both the $D$
and the $\phi$, we obtain an excess of 10.2$\pm$4.0 events in the 
$\phi$ signal region.

\begin{figure}
\center{\mbox{\psfig{figure=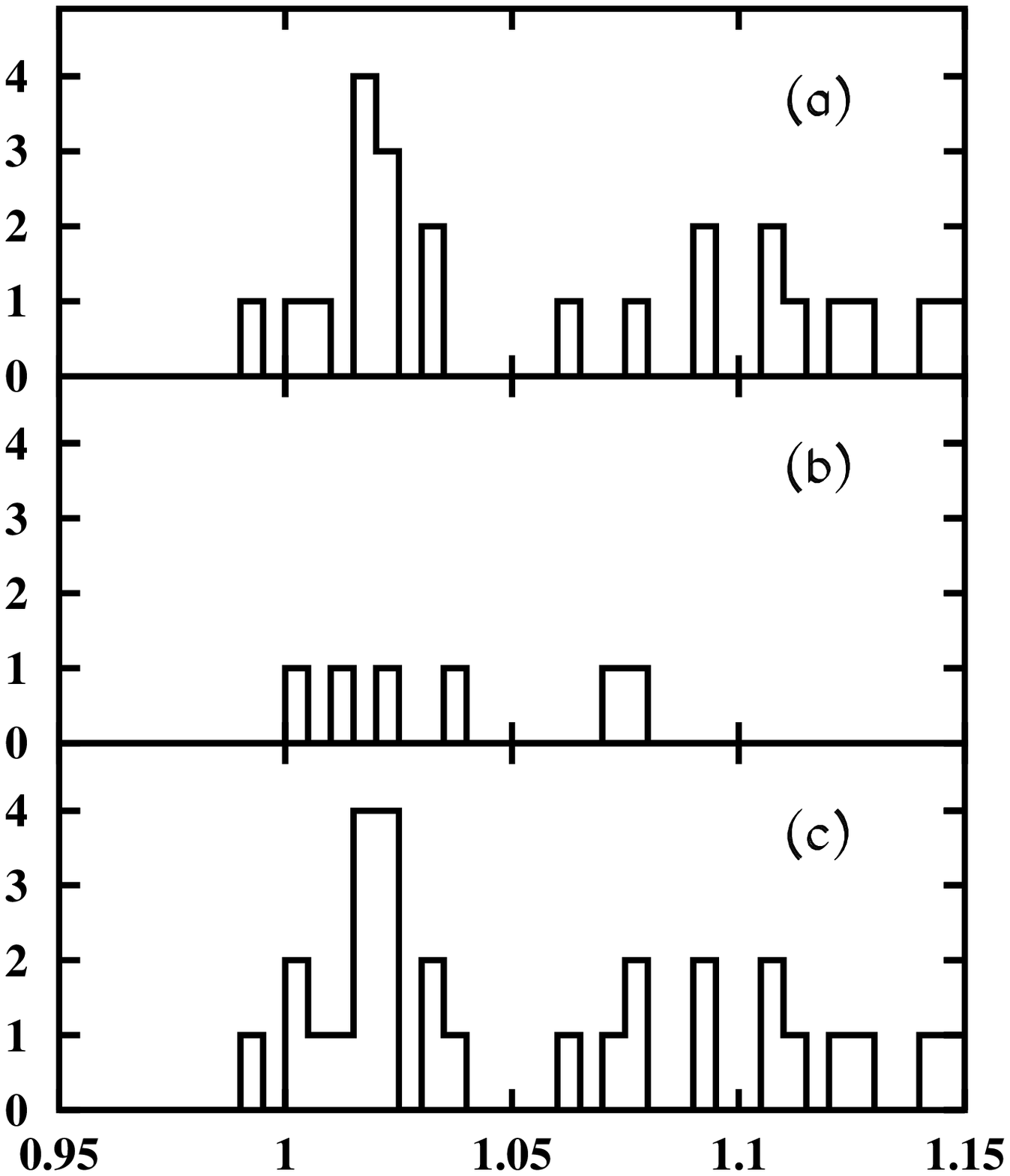,height=7.0cm,width=8.0cm}}}
\put(-165,6){\bf\large Inv. Mass (GeV/$c^2$)}
\put(-220,70){\bf\rotate{\large Events / 5 MeV/$c^2$}}
\vspace*{0.1cm}
\caption{Invariant mass distributions of ${K^+K^-}$ pairs
recoiling against fully reconstructed (a) $D^0$, (b) $D^+$,
and (c) $D^0$ or $D^+$.}
\end{figure}
\begin{figure}
\center{\mbox{\psfig{figure=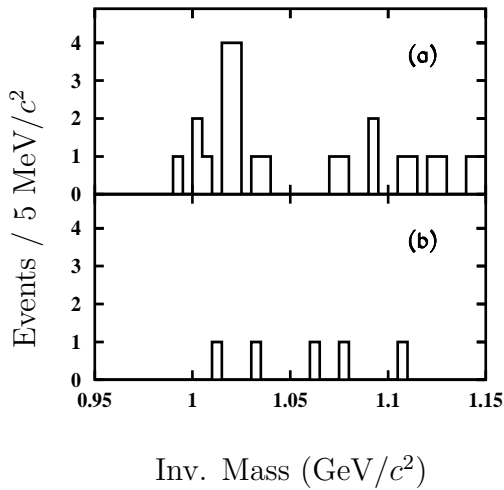,height=6.7cm,width=8.0cm}}}
\put(-165,0){\bf\large Inv. Mass (GeV/$c^2$)}
\put(-220,50){\bf\rotate{\large Events / 5 MeV/$c^2$}}
\vspace*{0.1cm}
\caption{Invariant mass distributions of ${K^+K^-}$ pairs
identified as from 
(a) ${D^0}$, and (b)  ${D^+}$.}
\end{figure}

Figures 6(a) and 6(b)
show the invariant mass of the ${K^+K^-}$
pairs from $D^+$ and $D^0$,
respectively, as identified by the
recoil charge criteria. 
After subtraction of backgrounds estimated using the $\phi$ and 
{\it D} side bands,
9.7$^{+4.3}_{-4.0}$ ${D^0\rightarrow\phi X^0}$ events
and 0.5$^{+1.9}_{-1.3}$
${D^+\rightarrow\phi X^+}$ events are determined
using Eqs. (3) and (4),
and a multinomial distribution that allows
the fluctuations of the signal and the background.
The branching fraction can be
obtained from the number of
{\it D} events ($N_{D}$)
in the recoil
against the tagged $D$ mesons, and the number of $D\phi$ events
($N_{D\phi }$):
\begin{equation}
B(D\rarr \phi X) = \frac {N_{D\phi}}
    {N_{D} \times \epsilon_{\phi}}.
\end{equation}
\noindent
where the $\phi$ detection efficiency
$\epsilon_{\phi}$=0.084$\pm$0.006 as
determined from Monte Carlo simulations.
Assuming 10.2$\pm$4.0 signal $\phi$ events, the average
branching fraction for the BES mixture of $D^0$ and $D^+$
mesons is measured to be
$$B({D} \rarr \phi X) = (1.34 \pm 0.52 \pm 0.12)\%,$$

\noindent
where the first error is statistical and second systematic. 

Based on 9.7$^{+4.3}_{-4.0}$  ${D^0\rightarrow\phi X^0}$ and
0.5$^{+1.9}_{-1.3}$  ${D^+\rightarrow\phi X^+}$
events, branching fraction  and 90\% C. L. limit
$$B({D^0 \rarr \phi X^0}) =(1.71^{+0.76}_{-0.71}\pm0.17)\%,$$
$$B({D^+} \rarr \phi X^+) <1.8\%, $$

\noindent
are obtained, where the first error is statistical
and the second systematic.
The systematic errors arise
from uncertainties
in the numbers of $D$ mesons and the  
error in the inclusive $\phi$ efficiency.
The final systematic errors are obtained by adding these
uncertainties in quadrature. 

\subsubsection{Search for the decay ${D^+}\rarr \phi e^+X$} 

Of the 15 $D\phi$ candidates selected
four are accompanied by at least one charged track within
$|\cos\theta|<0.85$.
These tracks are tested against electron hypothesis  where 
to identify electrons,
a confidence level of greater than $1\%$
is required, and ${L_e>L_{\pi}}$,
using \DEDX information. 
None of the tracks are identified as electrons.

\begin{small}
\begin{table}
\centering
\vspace{0.2cm}
\caption{Branching fractions and limits (90\% C. L.)} 
\begin{tabular}{|c|c|c|} 
decay mode & b. f. (\%)& Experiment  \\
\hline
${D^+\rightarrow \phi \pi^+}$& $0.61\pm 0.06$ &WA82 {\it et. al.}\\
\hline
$\phi \pi^+\pi^0$ & $2.3\pm 1.0$ & ACCMOR\\
\hline
$\phi K^+$        & $0.013^{+0.022}_{-0.019}$&E687 \\
\hline
$\phi e^+ \nu$    & $<2.09$ &MK3 \\
\hline
$\phi \mu^+\nu$   & $<3.72$ &MK3\\
\hline
$\phi\pi^+\pi^+\pi^-$ & $<0.2$ &E691  \\
\hline
${\phi e^+ X^0}$ & $<1.6$ & this experiment \\
${ \phi X^+}$        & $<1.8$ & this experiment\\
\hline
${D^0\rightarrow\phi \overline{K}^0}$  & $0.86\pm0.10$&ARGUS,CLEO, E687   \\ 
\hline
$\phi \pi^+ \pi^-$  & $0.108\pm0.029$ &ARGUS, CLEO, E687 \\ 
\hline
$\phi \pi^0$    & $<0.14$  &ARGUS\\ 
\hline
$\phi \eta$   & $<0.28$ &ARGUS\\
\hline
$\phi \omega$ & $<0.21$  & ARGUS\\ 
\hline
${ \phi X^0}$ & 1.71$^{+0.76}_{-0.71}\pm0.17$ &this experiment \\
\hline
${ D\rightarrow\phi X}$ & $1.34\pm0.52\pm0.12$ &this experiment  \\
\end{tabular}
\end{table}
\end{small}

From no observed ${D^+} \rarr \phi e^+ X^0$ event
in a sample of $2287$ $D^+$ decays,  and a
detection efficiency of 0.0652 for the decay,
a $90\%$ C.L. upper limit of
$B({D^+}\rarr \phi e^+ X^0)<1.6\%$ is placed.

\section{Discussion}

These BES results, together with branching fractions listed
in PDG98, are summarized in Table II.
Comparing with existing measurements
of exclusive $D^0$ and $D^+$ decays
containing a $\phi$ in the final states, as shown in Table II,
these BES branching fraction values indicate
little room for additional $\phi$ decay modes
of $D^0$ and $D^+$ mesons.

\section{Conclusion}
In summary, the absolute inclusive
branching fractions of the \dzero\ and \dplus\  mesons
decaying into a $\phi$ have been directly measured.
From a tagged sample of
9054$\pm$309$\pm$416  ${D\overline{D}}$ pairs,
10.2$\pm$4.0 $ {D}\rarr \phi X$
events are observed, leading to the first measurement of
$B( {D} \rarr \phi X) = (1.34 \pm 0.52 \pm 0.12)\%$
for a mixture of $D^0$ and $D^+$ mesons in the BES data sample,
$B( {D^0} \rarr \phi X^0) =(1.71^{+0.76}_{-0.71}\pm0.17)\%$,
$B( {D^+} \rarr \phi X^+) < 1.8\%$,  and
$B( {D^+}\rarr \phi e^+ X^0)<1.6\%$
at 90\% C. L. 

\section{Acknowledgements}
We would like to thank the staffs of the BEPC accelerator
and the Computing Center at the Institute of High Energy Physics,
Beijing, for their outstanding scientific efforts. The work of the BES
Collaboration was supported in part by
the National Natural
Science Foundation of China under Contract No. 19290400 and the Chinese
Academy of Sciences under contract No. H-10 and E-01 (IHEP),
and by the Department of
Energy under Contract Nos.
DE-FG03-92ER40701 (Caltech), DE-FG03-93ER40788 (Colorado State University),
DE-AC03-76SF00515 (SLAC), DE-FG03-91ER40679 (UC
Irvine), DE-FG03-94ER40833 (U Hawaii), DE-FG03-95ER40925 (UT Dallas).

\end{document}